\documentclass[a4paper,11pt]{article}
\usepackage{pos}
\usepackage{amsmath}

\title{Axion-Scalar Dynamics: from the Distance Conjecture to Cosmic Acceleration}

\author*[a]{Filippo Revello}

\affiliation[a]{Instituut voor Theoretische Fysica \& Leuven Gravity Institute, KU Leuven,\\ Celestijnenlaan 200D, B-3001 Leuven, Belgium}

\emailAdd{filippo.revello@kuleuven.be}

\abstract{We discuss the cosmology of axion-scalar systems in asymptotic limits of type IIB/F-theory flux compactifications. These results allow us to test a putative extension of the Distance Conjecture in a dynamical setting, which posits that towers of states should become exponentially light in the distance measured along the trajectory (as well as in the geodesic one). In the case of infinite distance limits, we review a known classification of late-time asymptotic solutions, which always verify the extension of the conjecture whenever all relevant effects are taken into account. We also extend the analysis to the case of finite distance limits, where the analogous statement would require trajectories approaching the singularity to have a finite length. Surprisingly, we find this is not the case for the class of models under consideration.  Moreover, the new solutions we find exhibit asymptotic accelerated expansion when approaching the boundary of moduli space.}

\FullConference{Proceedings of the Corfu Summer Institute 2025 "School and Workshops on Elementary Particle Physics and Gravity" (CORFU2025)\\
27 April - 28 September, 2025\\
Corfu, Greece\\
}

\begin{document}
\maketitle

\section{Introduction}

Cosmology represents one of our best hopes to connect String Theory with observations, given the very high energies reached in the earliest stages of the universe (see \cite{Cicoli:2023opf} for a review). Another aspect of early universe physics of particular interest to string theorists is the possibility of trans-Planckian displacements for scalar fields $\phi_i$, characterised by $\Delta \phi_i \gtrsim M_P$. On the one hand, large field displacements seem to be an important ingredient for inflationary models with a sizeable tensor-to-scalar ratio \cite{Baumann:2014nda}, which is an important goal for phenomenology. On the other hand, it is also clear that field displacements larger than $M_P$ can give rise to control issues in bottom-up Effective Field Theories (EFTs) with cutoff $\Lambda$ at or below the Planck scale. For large field displacements, a generic observable can be sensitive to the Wilson coefficients $c_n$ of the Planck-suppressed operators appearing in the EFT expansion, and thus require knowledge of the underlying UV completion. In the case of a single scalar $\phi$, for example, the potential can be expanded in the schematic form
\begin{equation}
    V(\phi)=\sum_{n \geq 2} c_n \left( \frac{\phi}{M_P}\right)^n,
\end{equation}
and all terms become important when $\Delta \phi \geq M_P$. While string theory should in principle be able to control such corrections, this cannot always be done within a \emph{single} EFT when traversing (super) Planckian distances. According to the  \emph{Swampland Distance Conjecture} \cite{Ooguri:2006in}, infinite distance limits in moduli space must always be accompanied by light towers of states - such as Kaluza-Klein (KK) or winding string modes - causing a breakdown of the EFT. In particular, whenever the \emph{geodesic distance} $d(P,Q)$ between two points $P$ and $Q$ approaches infinity, the mass of the lightest state is predicted to decay faster than
\begin{equation}\label{eq:mt}
    m_t \sim e^{-\alpha_d d(P,Q)},
\end{equation}
with $\alpha_d$ an $\mathcal{O}(1)$ coefficient that depends on the number of spacetime dimensions. As a side remark, let us also emphasize how the breakdown occurs whenever the lightest state in the tower becomes lower than the EFT cutoff - which is guaranteed to happen where the UV cutoff is time-independent. However, there do exist (rare) situations where the cutoff itself is time-dependent, and always stays below the (evolving) scale of the tower. An important example is given by \emph{volume kination}, corresponding to a universe dominated by the energy density of a volume modulus $\Phi$ that is rapidly rolling along a steep potential \cite{Conlon:2022pnx}. In that case, although a KK tower is becoming light in the large volume limit, the conditions $m_{\Phi}(t) \ll H(t) \ll m_{\rm{KK}}(t)$ are always verified.\footnote{Moreover, $\frac{{\rm{d}} m_{KK}}{{\rm{d}}t} \frac{1}{m_{KK}} \ll H$, which also guarantees the validity of the adiabatic approximation \cite{Baumann:2014nda}.}

The Distance Conjecture has been tested in a variety of String Theory settings where the low-energy EFT has an exact moduli space. Notable examples are 4d $\mathcal{N}=2$ theories resulting from type II compactifications on Calabi-Yau (CY) 3-folds \cite{Grimm:2018ohb}. However, it has thus far received less support in theories where the scalars acquire a potential \cite{Klaewer:2016kiy}, and there is no longer an exact moduli space. Unfortunately, this happens exactly for the $\mathcal{N}=1$ or $\mathcal{N}=0$ theories of interest to phenomenology. Another way to say this is that the distance conjecture, as originally stated, only applies to \emph{adiabatic} trajectories, which occur along geodesics in moduli space and where $\dot{\phi} \rightarrow 0$. On the other hand, in the presence of a non-zero potential, the trajectories will deviate from geodesics and acquire non-trivial spacetime dependence. One can see this from the equation of motion for the fields
\begin{equation}\label{eq:geod}
\ddot{\phi}^i+\Gamma_{j k}^i \dot{\phi}^j \dot{\phi}^k+(d-1) H \dot{\phi}^i+
\partial^i V=0,
\end{equation}
where Christoffel symbols are calculated (and indices raised) with the moduli space metric $G_{ij}(\phi_i)$. In the absence of a potential, \eqref{eq:geod} is exactly the geodesic equation on moduli space (up to non-affine reparametrisation), but this is no longer true when $\partial^i V \neq 0$. It is therefore natural to wonder whether an analogous statement to the distance conjecture will hold in these more realistic, dynamical scenarios. A minimal generalisation, first suggested in \cite{Landete:2018kqf}, is to conjecture that the tower of states predicted by the distance conjecture should obey\footnote{A possible counter-example was suggested in \cite{Buratti:2018xjt}, although strictly speaking it is not in a genuine EFT setting.}
\begin{equation}
m_t \sim e^{-\alpha_d \Delta_\gamma(P, Q)},
\end{equation}
where the \emph{dynamical distance} $\Delta_\gamma(P, Q)$ is given by the length of the trajectory followed between the points $P$ and $Q$,
\begin{equation}\label{eq:gdist}
\Delta_\gamma(P, Q)=\int_\gamma \mathrm{d} \tau \sqrt{G_{i j} \dot{\phi}^i \dot{\phi}^j}.
\end{equation}
In this paper, we will investigate the validity of this statement for asymptotic limits in type IIB String Theory/F-theory where a single complex-structure modulus is approaching the boundary of moduli space. Crucially, we will also consider the dynamics of the associated axion, corresponding to the imaginary part of the same $\mathcal{N}=1$ chiral multiplet where the geometric modulus resides. This will provide a setting that is simple enough to be treated analytically, yet rich enough in terms of dynamical features. 

As another application of our results, we will also highlight possible connections to phenomenology, and in particular the possibility of asymptotic accelerated expansion. The latter could be relevant to describe the current-day accelerated expansion in terms of quintessence or rolling scalar fields (see \cite{Andriot:2026lac} for a review). Unfortunately, it is notoriously difficult to embed a prolonged period\footnote{A short epoch of accelerated expansion can always be arranged near the inversion point of a scalar field that is climbing up some arbitrary potential.} of accelerated expansion in a controlled regime of string theory, as the required scalar field potentials typically end up being too steep. For a set of scalar fields in $d$ spacetime dimensions and with a potential $V(\phi_i)$, the strong form of the De Sitter conjecture postulates that \cite{Rudelius:2021oaz}
\begin{equation}\label{eq:sdS}
\gamma \equiv \frac{\|\nabla V(\varphi)\|}{V(\varphi)} \geq \frac{2}{\sqrt{d-2}},
\end{equation}
in the asymptotic region of moduli space. On the other hand, asymptotic accelerated expansion would require the complementary condition $\gamma \leq 2/ \sqrt{d-2}$. \footnote{In \cite{Calderon-Infante:2022nxb}, it was suggested that such a bound could be evaded in certain asymptotic limits involving multiple complex structure moduli in type IIB flux compactifications. However, these examples also require K\"ahler moduli stabilisation in order to fully work.} An interesting observation, first made in \cite{Cicoli:2020cfj,Cicoli:2020noz,Brinkmann:2022oxy}, is that the bound \eqref{eq:sdS} can be compatible with (asymptotic) accelerated expansion in curved moduli space. An axion and a saxion residing in the same multiplet are a typical example, and for this reason we also expect to find similar solutions in this setting.

Except for the finite distance limits, discussed in sections \ref{ssc:fd} and \ref{sc:imp}, the original results in this paper have been obtained in \cite{Grimm:2025cpq}, which we refer to for a more detailed exposition.

\section{Cosmology of the scalar-axion system}

As a concrete setting, let us consider two scalars $s,a$ described by the action
\begin{equation}
S=M_{P, d}^2 \int d^d x \sqrt{-g}\left\{\frac{1}{2} \mathcal{R}-\frac{1}{2} G_{s s} \partial_\mu s \partial^\mu s-\frac{1}{2} G_{a a} \partial_\mu a \partial^\mu a-V\right\}.
\end{equation}
In practice, we will identify $s$ with a geometric, complex-structure modulus and $a$ as the (pseudo-scalar) axion residing in the same $4d, \mathcal{N}=1$ chiral multiplet, $\Phi= s + ia$. As a consequence of the axionic shift symmetry when $V=0$, the components of the metric $G$ on moduli space will depend only on $s$, $G_{ss} \equiv G_{ss}(s)$ and $G_{aa}\equiv G_{aa}(s)$. On the other hand, the scalar potential will depend on both, $V \equiv V(s,a)$. 

In typical string compactifications, including the examples of section \ref{sec:ex}, the kinetic terms descend from a K\"ahler potential of the form $K=-2C \log\left( \Phi+ \bar{\Phi} \right)$, with $C>0$, and which results in the metric \footnote{We employ the convention $G_{I \bar{J}}= 2 \partial_{I} \partial_{\bar{J}} K$.}
\begin{equation}\label{eq:logm}
    G_{ss} = G_{aa}=\frac{C}{s^2}.
\end{equation}
With this form of the metric, the equations of motion \eqref{eq:geod} for the fields, supplemented by the first Friedmann equation, give rise to the system
\begin{equation}
\left\{\begin{array}{l}\label{eq:eoms}
\ddot{a}-\frac{2 \dot{a} \dot{s}}{s} + (d-1) H \dot{a}+ \frac{s^2}{C}\partial_a V=0 \\
\ddot{s}-\frac{\dot{s}^2}{s}+\frac{\dot{a}^2}{s} + (d-1) H \dot{s}+  \frac{s^2}{C} \partial_s V=0 \\
(d-1)(d-2)H^2=  C \frac{\dot{s}^2+\dot{a}^2}{s^2} +2V.
\end{array}\right.
\end{equation}
On the other hand, we will also encounter examples where the leading contribution to the K\"ahler potential vanishes and $C=0$. To analyse such cases, it is useful to recall the equations of motion for a more general form of the moduli space metric \cite{Revello:2023hro}, with $G_{ss}(s)=G_{aa}\equiv G(s)$:
\begin{equation}
\left\{\begin{array}{l}\label{eq:eomsG}
 \ddot{a}+\frac{G'(s)}{G(s)}\dot{a} \dot{s} + (d-1) H \dot{a}+ \frac{\partial_a V}{G(s)}=0 \\
\ddot{s}+\frac{G'(s)}{2G(s)} \left(\dot{s}^2-\dot{a}^2\right) + (d-1) H \dot{s}+  \frac{\partial_s V}{G(s)} =0\\
 (d-1)(d-2)H^2=  G(s) \left(\dot{s}^2+\dot{a}^2\right) +2V.
\end{array}\right.
\end{equation}

Let us now turn to the question of what scalar potentials $V(s,a)$ can appear in the above equations of motion. Explicit expressions for a certain class of string theory compactifications will be presented in section \ref{sec:ex}, but we anticipate that within such a framework the scalar potential can always be written in the form
\begin{equation}\label{eq:Vtot}
V(s, a)=\frac{1}{s^\lambda} \sum_{n=0}^N \frac{1}{s^n} P_n\left(\frac{a}{s}\right) \equiv \tilde{V}(s) \sum_{n=0}^N \frac{1}{s^n} P_n\left(\frac{a}{s}\right),
\end{equation}
where the $P_n(x)$'s are polynomials of degree bounded by $4$. On physical grounds, we also require
\begin{equation}\label{eq:Vs}
    V(s,a) \rightarrow 0 \qquad \text{as} \qquad s\rightarrow \infty,
\end{equation}
with $a$ fixed.

\section{Explicit examples from F-theory compactifications}\label{sec:ex}

We now present concrete examples of string theory embeddings for the axion-scalar systems considered in the previous section. In particular, we consider F-theory compactifications on elliptically fibred Calabi-Yau fourfolds with $4-$form flux.\footnote{These also include type IIB compactifications with 3-form fluxes $F_3$ and $H_3$.} They give rise to 4d $\mathcal{N}=1$ theories, whose scalar sector can be fully described (at the two-derivative level) by a K\"ahler potential $K$ and superpotential $W$. For chiral multiplets $\Phi_i=s_i+ia_i$, the corresponding action reads
\begin{equation}
S=M_P^2 \int d^4 x \sqrt{-g}\left[\frac{1}{2} \mathcal{R}-K_{I \bar{J}} \partial_\mu \Phi^I \partial^\mu \bar{\Phi}^{\bar{J}}-V(\Phi, \bar{\Phi})\right],
\end{equation}
where\footnote{In the following, the sum over K\"ahler moduli cancels out with the $-3|W|^2$ term, so that the potentials are positive definite for the complex-structure sector.}
\begin{equation}
V(\Phi, \bar{\Phi})=e^K\left(\sum_{\Phi_I, \Phi_J} K^{I \bar{J}} D_I W D_{\bar{J}} \bar{W}-3|W|^2\right).
\end{equation}
A complete classification of all asymptotic scalar potentials for a single complex structure modulus ($h^{3,1}=1$) was presented in \cite{Grimm:2025cpq}, using the tools of asymptotic Hodge Theory \cite{Grimm:2019ixq} (see \cite{vandeHeisteeg:2022gsp} for a review). An outcome of the analysis is that a finite number of different asymptotic limits can arise, depending on the singularity type, which can be located at both finite and infinite distance. They are denoted by what is known as the limit type, which corresponds to the mixed Hodge structure arising at the boundary. From this knowledge, it is possible to compute the (asymptotic) periods of the $(4,0)$ form $\Omega$, which in turn allow to reconstruct the physically relevant $K$ and $W$. We now list all corresponding flux potentials as a function of the $G_4$ fluxes $\left\{g_i \right\}$, where we have switched off the fluxes that give rise to terms not respecting \eqref{eq:Vs}. 

\paragraph{Type $\mathrm{I}_{0,1}$:}\hspace{-0.3cm}this is a finite distance limit, since the leading logarithmic contribution to the K\"ahler potential vanishes and
\begin{equation}
   K = - \log\big( 2 - 4 |A|^2  e^{-4\pi s} s \big),
\end{equation}
for some constant $A$. The scalar potential is
\begin{equation}
V=\frac{g_5^2+g_6^2}{s}.
\end{equation}

\paragraph{Type $\mathrm{I}_{1,1}$:}\hspace{-0.3cm}again we have a finite distance limit, where $K$ now takes the form
\begin{equation}
K = - \log\big(2-2 |A|^2 e^{-4 \pi  s} s^2 \big)\,,
\end{equation}
and the scalar potential is
\begin{equation}
V= \frac{g_5^2}{s^2}.
\end{equation}

\paragraph{Type $\mathrm{II}_{0,0}$:}\hspace{-0.3cm}this is an infinite distance limit, with
\begin{equation}
K = - \log\left(4s\right)\, ,
\end{equation}
and a potential of the form
\begin{equation}
    V=\frac{g_3^2+g_4^2}{s}.
\end{equation}

\paragraph{Type $\mathrm{III}_{0,0}$:}\hspace{-0.3cm}this is also infinite distance, with
\begin{equation}
    K = - \log\big(4s^2 \big)\, ,
\end{equation}
and
\begin{equation}
    V=\frac{2 \left(\left(g_5-a g_3\right){}^2+\left(g_6-a g_4\right){}^2\right)}{s^2}.
\end{equation}

\paragraph{Type $\mathrm{V}_{1,1}$:}\hspace{-0.3cm}it corresponds to the well-known Large Complex Structure (LCS) limit, which is located at infinite distance. Its data is given by
\begin{equation}
    K =- \log \big( \tfrac{2}{3}s^4 \big)\,,
\end{equation}
and
\begin{equation}
    V=\frac{6 \left(\left(a^2 g_3+2 \left(a g_4+g_5\right)\right){}^2+\left(a g_3+g_4\right){}^2\right)}{s^2}.
\end{equation}

The above examples will form the basis of our analysis. We notice how the K\"ahler potentials for the infinite distance limits all give rise to a metric of the form \eqref{eq:logm}, and a potential of the form \eqref{eq:Vtot}. For the finite distance limits, $C=0$ and exponential corrections become relevant. While this will complicate the dynamical system analysis, we also notice how for both finite distance limit types, the scalar potential satisfies
\begin{equation}\label{eq:Vexp}
    \partial_a V=0, \qquad \quad V(s)=\frac{g^2}{s^n}, \qquad \quad \frac{\partial_s V}{V}= -\frac{n}{s}, 
\end{equation}
where $n=1,2$ corresponds to types  $\mathrm{I}_{0,1}$ and  $\mathrm{I}_{1,1}$ respectively.
For the metric,
\begin{equation}\label{eq:Gexp}
   G(s) = A^2 s^n e^{-4 \pi s} \left(1+ \mathcal{O}\left( \frac{1}{s}\right) \right),  \qquad \frac{\partial_s  G(s)}{G(s)}= -4 \pi+ \mathcal{O}\left( \frac{1}{s}\right),
\end{equation}
where the subleading terms can be neglected in a large $s$ expansion.

\section{A dynamical systems analysis}

The goal of this section is to outline how one can classify the late-time solutions to the equations of motion \eqref{eq:eoms} and \eqref{eq:eomsG} using techniques from the theory of dynamical systems, in the spirit of \cite{Copeland:1997et}. 
\subsection{Infinite distance limits - single polynomial potential}\label{ssc:sc}
For simplicity of exposition, we first consider the simplified case where a single polynomial appears in \eqref{eq:Vtot}, namely
\begin{equation}
V(s, a)=\frac{1}{s^\lambda} P\left(\frac{a}{s}\right) \equiv \tilde{V}(s) P\left(\frac{a}{s}\right).
\end{equation}
One can introduce the variables
\begin{equation}
x=\frac{\dot{s}}{\alpha H s} \quad \quad y=\frac{\dot{a}}{\alpha H s} \quad \quad \quad z=\frac{1}{H} \sqrt{\frac{2 \tilde{V}(s)}{(d-1)(d-2)}} \quad \quad w=\frac{a}{s},
\end{equation}
where $\alpha=\sqrt{\frac{(d-1)(d-2)}{C}}$ and which satisfy the constraint
\begin{equation}\label{eq:cstr1}
x^2+y^2+z^2 P(w)=1.
\end{equation}
Then, the equations of motion \eqref{eq:eoms} can be recast as the autonomous, dynamical system
\begin{equation}
\left\{\begin{array}{l}\label{eq:xy}
\frac{d x}{d N}=-\alpha y^2-\left(1-x^2-y^2\right)\left[(d-1) x-\frac{\alpha}{2}\left(\lambda+\frac{w \partial_w P(w)}{P(w)}\right)\right] \\
\frac{d y}{d N}=\alpha x y-\left(1-x^2-y^2\right)\left[(d-1) y+\frac{\alpha}{2} \frac{\partial_w P(w)}{P(w)}\right] \\
\frac{d w}{d N}=\alpha(y-w x).
\end{array}\right.
\end{equation}
It is also instructive to introduce the new variables
\begin{equation}
S \equiv x^2+y^2, \quad \quad \quad T \equiv x+y w, 
\end{equation}
the first of which describes the total kinetic energy (scalar + axion) of the system. In terms of $S$ and $T$, the equations can be rewritten as
\begin{equation}
\left\{\begin{array}{l}\label{eq:ST}
\frac{d T}{d N}  =-(1-S)(d-1)\left[T-T_\lambda\right] \\
 \frac{d S}{d N}  =-(1-S)\left[2(d-1) S+\frac{d w}{d N} \frac{\partial_w P(w)}{P(w)}-\alpha \lambda x\right] \\
w'^2  \equiv\left(\frac{d w}{d N}\right)^2=\alpha^2\left[S\left(1+w^2\right)-T^2\right],
\end{array}\right.
\end{equation}
where $T_\lambda \equiv \frac{\alpha \lambda}{2(d-1)}$. A key observation is that the equation of motion for $T$ is now quite simple, and implies that $T \rightarrow T_{\lambda}$ asymptotically. This can be rigorously justified using the theory of Lyapunov functions \cite{Wiggins:2003}. The latter can be defined as a positive definite, monotonically decreasing functional $\mathcal{L}$ which vanishes at a single point in phase space, identified as the attractor. In this case, $\mathcal{L}=(T-T_{\lambda})^2$. From the second equation in \eqref{eq:ST}, it is possible to deduce that only two possibilities can arise for the variable $w$. The first one consists of a usual fixed point, where both $w$ and $S$ converge to a constant value
\begin{equation}
w \longrightarrow \bar{w} \quad \quad \quad  \quad S \longrightarrow \bar{S}=T_\lambda^2 /\left(1+\bar{w}^2\right).
\end{equation}
The value of $\bar{w}$ can be determined by requiring the RHS of \eqref{eq:ST} to vanish. As an aside, let us remark how the asymptotic limit for $T$ can be suggestively reformulated as
\begin{equation}
   2 G_{I \bar{J}}  \frac{{\rm d}}{{\rm d}N} \left( \Phi^I \bar{\Phi}^{\bar{J}}\right) = \frac{C}{s^2} \frac{{\rm d}}{{\rm d}N} \left( s^2+a^2\right) \rightarrow \frac{\lambda (d-2) }{2} \,.
\end{equation}

The other possibility, which can only be realised when there exists $w_0$ such that $P(w_0)=0$, is more interesting. In particular, it corresponds to the case where
\begin{equation}
    w \longrightarrow w_0 \quad \quad \quad \quad P(w) \rightarrow 0.
\end{equation}
In particular, the point $w=w_0$ is outside the domain where the system \eqref{eq:xy} is defined, as it leads to a divergence. Therefore, it does not give rise to a fixed point solution, but rather one where the remaining variables oscillate indefinitely.
Such oscillating solutions and their phenomenological properties will be the subject of the next section.

\subsection{Infinite distance limits - general potential}\label{ssc:gc}

The analysis of the general case, where the potential is of the form \eqref{eq:Vtot}, presents some significant complications. We will only state a few facts relevant for the analysis in the following sections, and refer the reader to \cite{Grimm:2025cpq} for a complete analysis. Upon defining the new variable $v=1/s$, the dynamical system becomes
\begin{equation}
\left\{\begin{array}{l}
\frac{d x}{d N}=-\alpha y^2-\left(1-x^2-y^2\right)\left[(d-1) x-\frac{\alpha}{2}\left(\lambda+\frac{\sum_{n=0}^N v^n\left(w \partial_w P_n(w)+n P_n(w)\right)}{\sum_{n=0}^N v^n P_n(w)}\right)\right] \\
\frac{d y}{d N}=+\alpha x y-\left(1-x^2-y^2\right)\left[(d-1) y+\frac{\alpha}{2} \frac{\sum_{n=0}^N v^n \partial_w P_n(w)}{\sum_{n=0}^N v^n P_n(w)}\right] \\
\frac{d w}{d N}=\alpha(y-w x) \\
\frac{d v}{d N}=-\alpha v x .
\end{array}\right.
\end{equation}

The most important outcome of the analysis is that, under very mild conditions, the oscillating solutions can no longer arise. Even if $P_0(w_0)=0$, if there exists $m>0$
such that $P_m(w_0) >0$, the late-time solution will converge to\footnote{Excluding the possibility $S\rightarrow 1$, see \cite{Grimm:2025cpq}.}
\begin{equation}
T \rightarrow T_{\lambda+m} \equiv \frac{\alpha(\lambda+m)}{2(d-1)} \quad \quad \quad \quad S \rightarrow \frac{T_{\lambda+m}^2}{1+w_0^2}.
\end{equation}
Intuitively, this is because when $P_0(w) \rightarrow 0$, sub-leading terms in the potential will dominate, and avoid the divergence present in the system \eqref{eq:xy}.

\subsection{Finite distance limits}\label{ssc:fd}

In the case of finite distance limits, $C=0$ and one must solve the more general system \eqref{eq:eomsG}. In particular, we can introduce new variables
\begin{equation}
    \tilde{x}= \frac{G(s) \dot{s}}{\tilde{\alpha} H} \qquad \quad \tilde{y}= \frac{G(s) \dot{a}}{\tilde{\alpha} H} \qquad \quad  w=\frac{2V(s)}{\tilde{\alpha}^2 H^2} ,
\end{equation}
where $\tilde{\alpha}=\sqrt{(d-1)(d-2)}$ and the tilde symbol is there to remind us that $\tilde{x},\tilde{y}$ \emph{do not} reduce to $x,y$ for $G(s)=\frac{C}{s^2}$. Indeed, they satisfy a constraint equation different from \eqref{eq:cstr1},
\begin{equation}\label{eq:cstr2}
    \tilde{x}^2+\tilde{y}^2+G(s)w=G(s),
\end{equation}
where $0\leq w \leq 1$. The system \eqref{eq:eomsG} is difficult to solve for arbitrary $G(s)$ and $V(s,a)$. However, for the specific form of the metric and scalar potential in the type $\mathrm{I}$ limits of section \ref{sec:ex}, one can use \eqref{eq:Vexp} and \eqref{eq:Gexp} to rewrite it in the more tractable form\footnote{Sub-leading corrections of order $\mathcal{O}\big(\frac{1}{s}\big)$ in \eqref{eq:Gexp} have been consistently neglected, assuming the evolution takes place in the asymptotic region $s \gg 1$. } 
\begin{equation}
\left\{\begin{array}{l}\label{eq:xyf}
\frac{d \tilde{x}}{d N}=-2 \pi \tilde{\alpha}(1-w)-(d-1)\tilde{x} w +\frac{\tilde{\alpha}n}{2s} w\\
\frac{d \tilde{y}}{d N}= - (d-1)\tilde{y} w \\
\frac{d w}{d N}=2w(1-w) \left[(d-1)- \frac{\tilde{\alpha}n}{2s}  \frac{ \tilde{x}}{\tilde{x}^2+\tilde{y}^2}\right]
\end{array}\right.
\end{equation}
As long as $G(s)$ is invertible, \eqref{eq:xyf} is an autonomous dynamical system, as $s$ depends on the other coordinates implicitly through \eqref{eq:cstr2}. The equation for $\tilde{y}$ is particularly simple, and implies that $\tilde{y}^2$ is positive and monotonically decreasing. Therefore, $\tilde{y}$ will converge to a constant as $N \rightarrow \infty$, $\tilde{y} \rightarrow \tilde{y}_{\infty}$. If $\tilde{y}_{\infty} \neq 0$, one can integrate the equation of motion for $\tilde{y}$ to show that the improper integral
\begin{equation}\label{eq:int1}
  I_1 \equiv  \int_{N_0}^{\infty} {\rm d} N' w(N') 
\end{equation}
converges, where $N_0$ is some appropriate initial time. In turn, since $|w'|$ is bounded as $N\rightarrow \infty$,\footnote{The non-zero limit of $\tilde{y}_{\infty} $ is crucial here.} $w(N)$ is uniformly continuous and \emph{Barbalat's Lemma} implies that $w \rightarrow 0$. However, this is incompatible with the differential equation for $w$, since $w$ cannot converge to zero if $\frac{d w}{d N} >0$ as $N\rightarrow \infty$. We must conclude that $\tilde{y} \rightarrow 0$ as $N \rightarrow \infty$, and the integral $I_1$ in \eqref{eq:int1} diverges. 

We now restrict to the branch of solutions that approaches the finite-distance singularity where $G(s) \rightarrow 0$ as $s\rightarrow \infty$.\footnote{It is quite intuitive that the evolution will drive the system down the potential and towards large $s$. Moreover, there exist no stable fixed points of \eqref{eq:xyf} with $\tilde{x}\neq 0$.}
On this branch, the constraint \eqref{eq:cstr2} already tells us that $\tilde{x},\tilde{y} \rightarrow 0$. Integrating the first equation in \eqref{eq:xyf} between $N$ and $N+1$, for a given $\varepsilon$ there exists $\tilde{N}$ such that
\begin{equation}
    I_2(N)\equiv 2 \pi \tilde{\alpha} \int_{N}^{N+1} {\rm d} N'\left(1- w(N') \right)< |\tilde{x}(N+1)-\tilde{x}(N)|+ \varepsilon
\end{equation}
for any $N>\tilde{N}$, since $\tilde{x} \rightarrow 0$ and $s \rightarrow \infty$. Then, $I_2(N) \rightarrow 0$ as $N \rightarrow \infty$. Since $0 \leq w \leq 1$ and $w'$ is bounded from above, arbitrarily narrow spikes are excluded and $w \rightarrow 1$. In this regime, the asymptotic solution reads
\begin{equation}\label{eq:sfd}
    \tilde{x}(N) = k_1^2 \frac{(d-1)^2}{\pi \tilde{\alpha} n} N^{n+1}e^{-2(d-1)N} \quad \tilde{y}(N)=k_1 N^{n/2}e^{-(d-1)N} \quad w(N)=1-\frac{n}{2(d-1)N},
\end{equation}
where we have omitted terms of relative order $\mathcal{O}\left( \frac{\log (N)}{N}\right)$ and higher, and $k_1$ is an integration constant. One can also express $s(N)$ as
\begin{equation}
  s(N)=  \frac{d-1}{2 \pi} N-\frac{1}{4 \pi} \log (N)+ s_0,
\end{equation}
with $s_0$ another integration constant. Surprisingly, we find that the ratio $\tilde{y}/\tilde{x}$ is not bounded as $N\rightarrow \infty$, but rather it diverges exponentially. Implications for the dynamical distance conjecture will be discussed shortly.

From a phenomenological perspective, the acceleration parameter is given by
\begin{equation}\label{eq:aex}
    \varepsilon \equiv -\frac{\dot{H}}{H^2}=(d-1)(1-w) \simeq \frac{n}{2N} \quad \text{as} \quad N \rightarrow \infty,
\end{equation}
giving rise to an asymptotic phase of accelerated expansion. As mentioned in the introduction, this is a remarkable property, that is extremely hard to realise in top-down string theory examples. To obtain an intuitive understanding of how an accelerating phase takes place, we can compare our scenario to the one suggested in \cite{Cicoli:2020cfj,Cicoli:2020noz,Brinkmann:2022oxy,Revello:2023hro}. Both cases feature an identical runaway potential that only depends on the saxion field $s$ as in \eqref{eq:Vexp}, but the moduli space metrics are different. For the multi-field solutions \cite{Cicoli:2020cfj,Cicoli:2020noz,Brinkmann:2022oxy,Revello:2023hro}, $G(s)$ is given by \eqref{eq:logm}, and has constant curvature proportional to $1/C$. Accelerated expansion only arises if $C$ is small enough, and in the limit $C \rightarrow 0$ one has $ \varepsilon \rightarrow 0 $. Therefore, the accelerated expansion in \eqref{eq:aex} can be understood as the effect of the diverging moduli space curvature for the metric \eqref{eq:Gexp} as $s \rightarrow \infty$. So far, we have neglected the effect of K\"ahler moduli, whose presence would induce a new runaway direction and potentially spoil the accelerating phase. A more thorough analysis of this scenario and its relevance to real-world cosmology is therefore left to future work.

\section{Implications for the Distance Conjecture}\label{sc:imp}

The classification of the previous section can be used to infer physical information on the asymptotic trajectories followed by the scalar fields. In particular, we would like to understand whether the tower of states predicted by Eq. \eqref{eq:mt} also becomes exponentially light in terms of the dynamical distance $\Delta(p,q)$. This is equivalent to requiring that the dynamical distance $\Delta(p,q)$ does not grow more than linearly as a function of the geodesic distance $d(p,q)$. From \eqref{eq:gdist}, one can express
\begin{equation}\label{eq:d1}
    \Delta_{\gamma}(P,Q) =  \int_{s(P)}^{s(Q)} {\rm{d}} s \sqrt{G(s)} \sqrt{1+\left( \frac{{\rm{d}}a}{{\rm{d}}s}\right)^2 }.
\end{equation}
Therefore, the \emph{dynamical} form of the distance conjecture is verified whenever the ratio $\left| \frac{{\rm{d}}a}{{\rm{d}}s}\right|= \left| \frac{y}{x}\right|$ is bounded. Conversely, if this ratio diverges sufficiently fast, the dynamical distance can grow faster than the geodesic one, and the criterion must be checked explicitly.\footnote{Oscillating cases also require a more careful evaluation of the integral.}

\begin{figure}[h!]
\centering
 \includegraphics[width=0.4\textwidth]{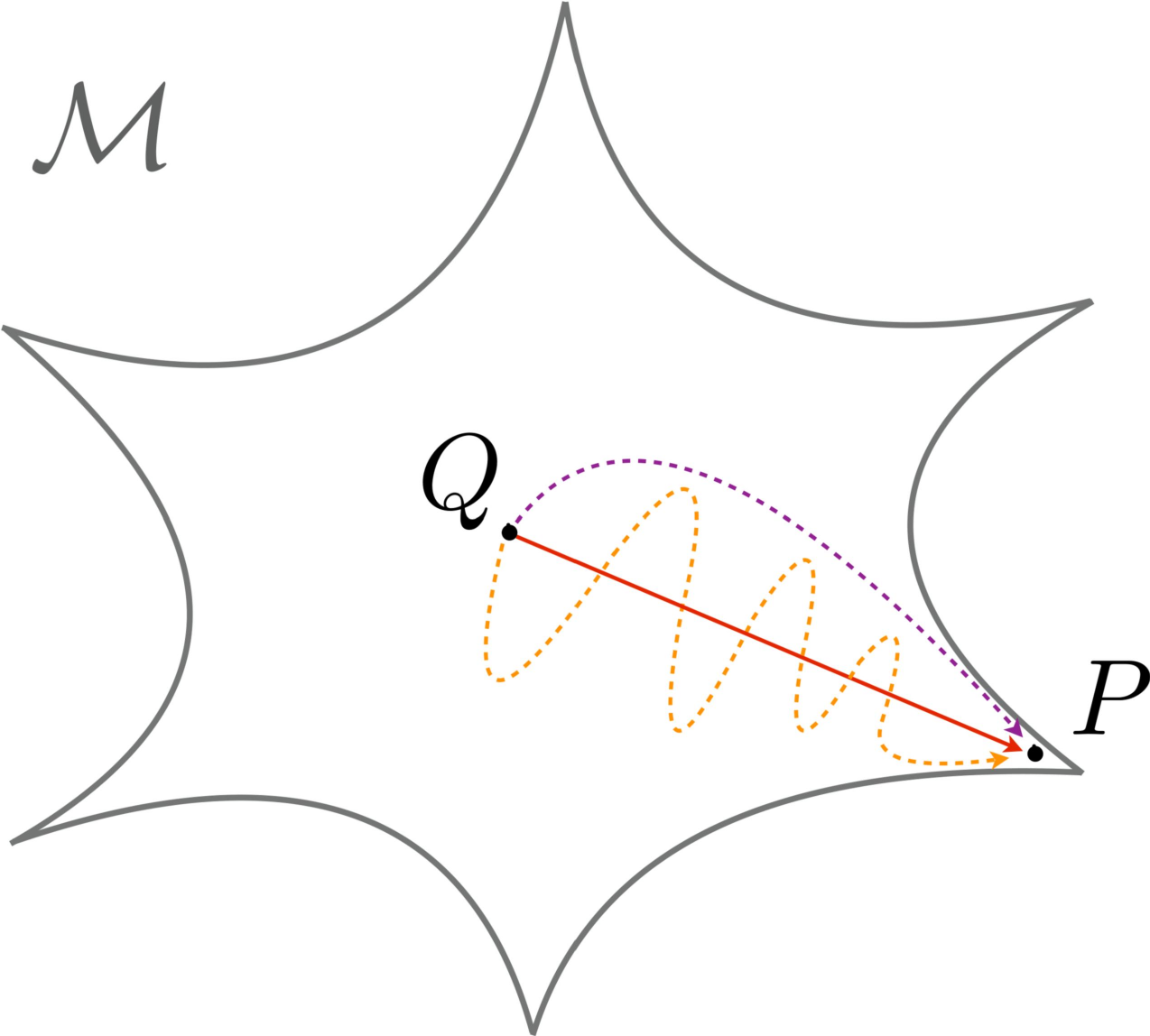}
\caption{Artistic depiction of two different trajectories approaching the boundary of moduli space, representing the geodesic distance $d(P,Q)$ and the dynamical one $\Delta(P,Q)$. Figure taken from \cite{Grimm:2025cpq}.}
\label{fig:traj}
\end{figure}

We can start by discussing the infinite distance limits, corresponding to the dynamical systems studied in subsections \ref{ssc:sc} and \ref{ssc:gc}. In terms of the $S,T$ variables, one can also write
\begin{equation}\label{eq:da/ds}
    \frac{{\rm d}a}{{\rm d}s} = \frac{y}{x}= \frac{T w \pm \sqrt{S(1+w^2)-T^2}}{T \mp w \sqrt{S(1+w^2)-T^2}}.
\end{equation}
If the evolution reaches a fixed point, the square root terms vanish asymptotically and $y/x \rightarrow \bar{w}$, and the dynamical form of the distance conjecture is satisfied.
On the other hand, for oscillatory solutions some more care must be taken, making sure that the denominator of \eqref{eq:da/ds} cannot vanish. As a reminder, such solutions are characterised by $w\rightarrow w_0$, with $P(w_0)=0$. In particular, this can happen if $w_0=0$ and $T_{\lambda}=0$. It turns out that an example of this kind can indeed arise in the classification of \cite{Grimm:2025cpq}. In the Large Complex Structure (LCS) limit, corresponding to a singularity of type $\mathrm{V}_{1,1}$ in the Hodge theory classification, a suitable combination of fluxes (denoted by $f$) generates a potential of the form
\begin{equation}
     P(w)= 3 f^2 w^2(1+w^2) \quad \quad \quad \lambda=0.
\end{equation}
Moreover, a similar potential can also arise in the case of a $\mathrm{III}_{0,0}$ singularity.\footnote{Unlike the LCS point, it is not known whether this case can explicitly arise in CY compactifications.}
Asymptotically, an approximate solution is given by
\begin{equation}\label{eq:eomsol}
    \begin{split}
    & x(N) =  \left[ k_3-\sqrt{\frac{2}{3}}  \sin \left(2 e^{ \frac{3}{2} (N-N_0)}  +2 k_4\right) \right]e^{-\frac{3}{2} (N-N_0)} \\ & y(N) = \cos \left( e^{ \frac{3}{2}(N-N_0)}+k_4 \right) \\ & w(N) =  \frac{4 \sqrt{6}}{3} e^{- \frac{3}{2} (N-N_0)} \sin \left( e^{\frac{3}{2} (N-N_0)} +k_4 \right),
    \end{split}
\end{equation}
with $k_3,k_4$ and $N_0$ constants. One can use \eqref{eq:eomsol} to infer that $s(N)$ reaches a finite value. The dynamical distance $\Delta(P,Q)$ does not, so it clearly diverges also as a function of the geodesic one $d(P,Q)$, leading to an apparent violation of the dynamical version of the distance conjecture.

However, as we will shortly see, one should not jump too hastily to negative conclusions, at least in this case. On physical grounds, the solution \eqref{eq:eomsol} at late times is essentially equivalent to a stabilised saxion and an axion oscillating around a quadratic potential - which could equally well arise in a stable Minkowski vacuum. In realistic physical scenarios, one would expect the oscillations to be damped exponentially in terms of the time $t$, leading to a finite distance $\Delta(P,Q)$. This is exactly what happens during reheating, where the role of friction is caused by decays of the oscillating field. Yet another possibility is that sub-leading corrections play a role, and eventually dominate the asymptotic solution as $P(w)\rightarrow 0$. In the LCS case, $\alpha'$ corrections to the K\"ahler potential indeed generate corrections to the potential of the form
\begin{equation}
    V(s,a) \supset   \frac{243 g_3^2 \xi^2}{4 s^6} \quad \quad \longrightarrow \quad \quad m=6, \quad \quad P_6(w) = \frac{243}{4}\xi^2 g_3^2,
\end{equation}
where $g_3$ is a flux parameter (as before) and $\xi$ depends on topological data of the 4-fold. According to the analysis sketched in \ref{ssc:gc}, this term will drive the evolution towards an actual fixed point, as it does not vanish when $w\rightarrow 0$. In \cite{Grimm:2025cpq}, this conclusion was also verified numerically.

Finally, let us move to finite distance limits, whose classical equations of motion were studied in subsection \ref{ssc:fd}. In both limit types, the scalar potential takes a very simple form in the classification of section \ref{sec:ex}, and the metric always satisfies \eqref{eq:Gexp}. Asymptotic solutions to the equations of motion were given in \eqref{eq:sfd}, from which one can easily see that
\begin{equation}
    \frac{\tilde{y}}{\tilde{x}} \sim N^{-n/2-1}e^{(d-1)N},
\end{equation}
violating the finite-distance analogue of the dynamical-distance criterion formulated above. In other words, there appear to exist trajectories approaching the finite distance boundary that have an infinite length. One could wonder whether subleading corrections to the potential might be able to modify the picture, as discussed previously for oscillating solutions. In this case, however, any corrections to the potential that are subleading in a $1/s$ will not have any impact on the dynamics, as they do not depend on $a$. In principle, exponentially small corrections to the potential (that may depend on $a$) could lead to a different asymptotic evolution, but the exponential suppression in $s$ makes it unlikely. In any case, the analysis is much more complicated, and deferred to future work. A last, possible way out would be for axion decays to play a role, but the axions are massless at this level.

\section{Conclusions}

We discussed a possible generalisation of the distance conjecture to a dynamical setting, according to which towers of states should become light in terms of the \emph{dynamical distance} (rather than the geodesic distance), defined as the length of the trajectory in moduli space. To do so, we leveraged the existing classification of all one-modulus asymptotic (complex structure) limits in type IIB/F-theory compactifications \cite{Grimm:2025cpq}. From a technical perspective, this relies on a dynamical systems analysis of the associated equations of motion, and it results in a complete classification of their asymptotic behaviour. Along the way, we uncovered the existence of peculiar \emph{oscillating} solutions, where the kinetic energy density of the axion oscillates indefinitely with a fixed amplitude. While such solutions would appear to violate the modified version of the distance conjecture suggested above, we also showed how sub-leading corrections can destabilize these oscillating solutions and restore the validity of the statement. All in all, the extension of the conjecture seems to hold in all asymptotic, one-modulus limits with an infinite distance singularity.

On the other hand, we also generalised the results of \cite{Grimm:2025cpq} to finite distance singularities, where the dynamical version of the distance conjecture does not seem to hold. In particular, for both limit types under consideration, the curves traced by the dynamical trajectories have infinite length, despite the singularity being located at finite distance. While one could always try to invoke unknown, sub-leading corrections that may eventually cure the pathological behaviour, the conclusion seems fairly robust in this case.

From a phenomenological perspective, the finite distance solutions within the truncated complex-structure sector realise asymptotic accelerated expansion at the boundary of moduli space, with an $\varepsilon$ parameter that approaches zero at infinity. Whether this survives after including K\"ahler moduli and other consistency conditions remains an important open question, to understand eventually whether realistic models of quintessence can be constructed in this way.\footnote{See \cite{Lanza:2024uis} for models aiming to achieve accelerated expansion in an intermediate region of moduli space located between bulk and boundary.} Moreover, the examples we have studied remind us of the dynamical relevance of axions in cosmological situations, both through their kinetic couplings to saxions \cite{Cicoli:2020cfj,Cicoli:2020noz,Brinkmann:2022oxy,Revello:2023hro} and in the scalar potential. Another avenue of future exploration would be to understand whether moduli- and axion-dominated solutions could have applications to describe the (largely unconstrained) epoch between the end of inflation and reheating \cite{Cicoli:2023opf,Apers:2024ffe}. In this context, we have seen how dynamical system techniques can be particularly useful to understand the asymptotic solutions to the background equations, even in very involved cases (see also \cite{Shiu:2023fhb,Shiu:2024sbe} for recent progress). It would be interesting to extend the reach of these techniques to even more complicated settings, such as those involving multi-moduli limits and scalar potentials involving axions.

From a long-term perspective, we advocate for a more careful consideration of dynamical effects in the study of Swampland conjectures, in order to connect more directly with realistic cosmological scenarios and constrain them more effectively. We also stress the importance of studying moduli space limits that do not necessarily fall under the lamppost of large volume or large complex structure, as they can sometimes lead to phenomenological surprises.

\subsection*{Acknowledgments}
The author gratefully acknowledges Thomas Grimm and Damian van de Heisteeg for discussions and the collaboration leading to \cite{Grimm:2025cpq}, on which these proceedings are based. I would also like to thank the organizers and participants of CORFU2025 for the stimulating atmosphere, and Flavio Tonioni for comments on the draft. The research of FR was supported by a junior postdoctoral fellowship of the Fonds Wetenschappelijk Onderzoek (FWO), project number 12A1Q25N.


\begin{thebibliography}{99}

\bibitem{Cicoli:2023opf}
M.~Cicoli, J.~P.~Conlon, A.~Maharana, S.~Parameswaran, F.~Quevedo and I.~Zavala,
``String cosmology: From the early universe to today,''
Phys. Rept. \textbf{1059} (2024), 1-155
doi:10.1016/j.physrep.2024.01.002
[arXiv:2303.04819 [hep-th]].

\bibitem{Baumann:2014nda}
D.~Baumann and L.~McAllister,
``Inflation and String Theory,''
Cambridge University Press, 2015,
ISBN 978-1-107-08969-3, 978-1-316-23718-2
doi:10.1017/CBO9781316105733
[arXiv:1404.2601 [hep-th]].

\bibitem{Ooguri:2006in}
H.~Ooguri and C.~Vafa,
``On the Geometry of the String Landscape and the Swampland,''
Nucl. Phys. B \textbf{766} (2007), 21-33
doi:10.1016/j.nuclphysb.2006.10.033
[arXiv:hep-th/0605264 [hep-th]].

\bibitem{Conlon:2022pnx}
J.~P.~Conlon and F.~Revello,
``Catch-me-if-you-can: the overshoot problem and the weak/inflation hierarchy,''
JHEP \textbf{11} (2022), 155
doi:10.1007/JHEP11(2022)155
[arXiv:2207.00567 [hep-th]].

\bibitem{Grimm:2018ohb}
T.~W.~Grimm, E.~Palti and I.~Valenzuela,
``Infinite Distances in Field Space and Massless Towers of States,''
JHEP \textbf{08} (2018), 143
doi:10.1007/JHEP08(2018)143
[arXiv:1802.08264 [hep-th]].

\bibitem{Klaewer:2016kiy}
D.~Klaewer and E.~Palti,
``Super-Planckian Spatial Field Variations and Quantum Gravity,''
JHEP \textbf{01} (2017), 088
doi:10.1007/JHEP01(2017)088
[arXiv:1610.00010 [hep-th]].

\bibitem{Landete:2018kqf}
A.~Landete and G.~Shiu,
``Mass Hierarchies and Dynamical Field Range,''
Phys. Rev. D \textbf{98} (2018) no.6, 066012
doi:10.1103/PhysRevD.98.066012
[arXiv:1806.01874 [hep-th]].

\bibitem{Rudelius:2021oaz}
T.~Rudelius,
``Dimensional reduction and (Anti) de Sitter bounds,''
JHEP \textbf{08} (2021), 041
doi:10.1007/JHEP08(2021)041
[arXiv:2101.11617 [hep-th]].

\bibitem{Buratti:2018xjt}
G.~Buratti, J.~Calder{\'o}n and A.~M.~Uranga,
``Transplanckian axion monodromy!?,''
JHEP \textbf{05} (2019), 176
doi:10.1007/JHEP05(2019)176
[arXiv:1812.05016 [hep-th]].

\bibitem{Grimm:2019ixq}
T.~W.~Grimm, C.~Li and I.~Valenzuela,
``Asymptotic Flux Compactifications and the Swampland,''
JHEP \textbf{06} (2020), 009
[erratum: JHEP \textbf{01} (2021), 007]
doi:10.1007/JHEP06(2020)009
[arXiv:1910.09549 [hep-th]].

\bibitem{vandeHeisteeg:2022gsp}
D.~T.~E.~van de Heisteeg,
``Asymptotic String Compactifications: Periods, flux potentials, and the swampland,''
doi:10.33540/1380
[arXiv:2207.00303 [hep-th]].

\bibitem{Grimm:2025cpq}
T.~W.~Grimm, D.~van de Heisteeg and F.~Revello,
``Axion-Scalar Systems and Dynamical Distances,''
[arXiv:2510.12879 [hep-th]].

\bibitem{Copeland:1997et}
E.~J.~Copeland, A.~R.~Liddle and D.~Wands,
``Exponential potentials and cosmological scaling solutions,''
Phys. Rev. D \textbf{57} (1998), 4686-4690
doi:10.1103/PhysRevD.57.4686
[arXiv:gr-qc/9711068 [gr-qc]].

\bibitem{Wiggins:2003}
S.~Wiggins,
``Introduction to Applied Nonlinear Dynamical Systems and Chaos'',
Springer, New York (2003),
ISBN 0387001778, 9780387001777,
doi:10.1007/b97481.

\bibitem{Lanza:2024uis}
S.~Lanza and A.~Westphal,
``Uplifts in the penumbra: features of the moduli potential away from infinite-distance boundaries,''
JHEP \textbf{05} (2025), 071
doi:10.1007/JHEP05(2025)071
[arXiv:2412.12253 [hep-th]].

\bibitem{Cicoli:2020cfj}
M.~Cicoli, G.~Dibitetto and F.~G.~Pedro,
``New accelerating solutions in late-time cosmology,''
Phys. Rev. D \textbf{101} (2020) no.10, 103524
doi:10.1103/PhysRevD.101.103524
[arXiv:2002.02695 [gr-qc]].

\bibitem{Cicoli:2020noz}
M.~Cicoli, G.~Dibitetto and F.~G.~Pedro,
``Out of the Swampland with Multifield Quintessence?,''
JHEP \textbf{10} (2020), 035
doi:10.1007/JHEP10(2020)035
[arXiv:2007.11011 [hep-th]].

\bibitem{Brinkmann:2022oxy}
M.~Brinkmann, M.~Cicoli, G.~Dibitetto and F.~G.~Pedro,
``Stringy multifield quintessence and the Swampland,''
JHEP \textbf{11} (2022), 044
doi:10.1007/JHEP11(2022)044
[arXiv:2206.10649 [hep-th]].


\bibitem{Calderon-Infante:2022nxb}
J.~Calder{\'o}n-Infante, I.~Ruiz and I.~Valenzuela,
``Asymptotic accelerated expansion in string theory and the Swampland,''
JHEP \textbf{06} (2023), 129
doi:10.1007/JHEP06(2023)129
[arXiv:2209.11821 [hep-th]].


\bibitem{Andriot:2026lac}
D.~Andriot,
``Dark energy from string theory: an introductory review,''
[arXiv:2603.25797 [hep-th]].

\bibitem{Revello:2023hro}
F.~Revello,
``Attractive (s)axions: cosmological trackers at the boundary of moduli space,''
JHEP \textbf{05} (2024), 037
doi:10.1007/JHEP05(2024)037
[arXiv:2311.12429 [hep-th]].

\bibitem{Apers:2024ffe}
F.~Apers, J.~P.~Conlon, E.~J.~Copeland, M.~Mosny and F.~Revello,
``String theory and the first half of the universe,''
JCAP \textbf{08} (2024), 018
doi:10.1088/1475-7516/2024/08/018
[arXiv:2401.04064 [hep-th]].

\bibitem{Shiu:2023fhb}
G.~Shiu, F.~Tonioni and H.~V.~Tran,
``Late-time attractors and cosmic acceleration,''
Phys. Rev. D \textbf{108} (2023) no.6, 063528
doi:10.1103/PhysRevD.108.063528
[arXiv:2306.07327 [hep-th]].

\bibitem{Shiu:2024sbe}
G.~Shiu, F.~Tonioni and H.~V.~Tran,
``Analytic bounds on late-time axion-scalar cosmologies,''
JHEP \textbf{09} (2024), 158
doi:10.1007/JHEP09(2024)158
[arXiv:2406.17030 [hep-th]].


\end{thebibliography}
\end{document}